\documentstyle[12pt]{article}
\begin{document}
\bibliographystyle{aip}

\begin{center}
{\bf X-RAY ABSORPTION STUDY OF PULSED LASER DEPOSITED BORON NITRIDE
FILMS}\\
\end{center}
A.  CHAIKEN$^*$, L.J. TERMINELLO$^*$, J. WONG$^*$, G.L. DOLL$^{\dag}$, AND T.
SATO$^{\ddag}$\\
$^*$Lawrence Livermore National Laboratory, Livermore, CA 94550\\
$^{\dag}$GM Research and Development Laboratory, Warren, MI 48090\\
$^{\ddag}$NIRIM, Tsukuba City, Japan

\begin{flushleft}
{\bf ABSTRACT}
\end{flushleft}
B and N K-edge x-ray absorption spectroscopy measurements have been
performed on three BN thin films grown on Si substrates using
ion-assisted pulsed laser deposition.    Comparison of the films'
spectra to those of several single-phase BN powder standards shows
that the films consist primarily of $sp^2$ bonds.    Other features in the
films' spectra suggest the presence of secondary phases, possibly
cubic or rhombohedral BN.    Films grown at higher deposition rates
and higher ion-beam voltages are found to be more disordered, in
agreement with previous work.

\begin{flushleft}
{\bf INTRODUCTION}
\end{flushleft}

Near-edge x-ray absorption spectroscopy is well-known as a method for
characterizing the bonding and orientation of organic molecules
adsorbed on thin-film surfaces.\cite{stohrbook} Subsequent work has
shown that core-level x-ray absorption is also a useful technique for
characterizing the unoccupied electronic states of low-atomic-number
solids.  A number of x-ray absorption studies have focussed on
graphite\cite{rosenberg86
} and diamond,\cite{morar85} the
two common crystalline phases of carbon.  Other researchers have
compared the spectra of amorphous carbon and hydrogenated diamond-like
carbon thin films to those of the bulk crystalline
phases.\cite{denley80}

Boron nitride is isoelectronic to carbon and has both hexagonal and
cubic phases analogous to graphite and diamond.
In hexagonal BN (hBN), B and N atoms in alternate layer planes lie
directly on top of one another in an AA$^{\prime}$A stacking
arrangement, as opposed to the staggered ABA stacking in graphite,
which causes C atoms in neighboring layers to be offset.  Rhombohedral
BN (rBN) differs from hBN only in its staggered ABCA stacking.
Previous NEXAFS studies have compared the B and N K-edge spectra of
well-ordered hexagonal, cubic, wurtzite and rhombohedral BN
powders.\cite{terminello93} The insight obtained through analysis of
the powder spectra can be used to interpret similar data taken on thin
BN films, which  are likely to find use as both tribological
coatings and as wide-gap semiconducting materials.

\begin{flushleft}
{\bf EXPERIMENTAL DETAILS}
\end{flushleft}

The three BN films used in this study were produced by ion-assisted
pulsed laser deposition onto heated silicon substrates, as detailed
previously.\cite{ballal92}    Deposition conditions are summarized in
Table 1.
Infrared transmission measurements (not shown) on
these films revealed absorption at wavelengths characteristic of both
the cubic and hexagonal phases, allowing an estimate of their volume
fraction.     The purity of the three types of BN powder was
determined using x-ray diffraction (not shown) to be greater than 90\%.

\samepage{
\begin{small}
\begin{table}[t]
\begin{center}
\begin{tabular}{||l|l|l|l||}
\hline
\hline
Peak Energy (eV) & Peak Separation (eV) & Rel. Integ. Intens. &
Lorentzian Width \\
\hline
\hline
\multicolumn{4}{||c||}{V$\rm _{ion}$ = 0.48 kV; Ar and
N ions; deposition rate 0.4 \AA/s; approx. 30 \% $sp^3$ and 70\% $sp^2$} \\
\hline
192.05 & NA & 1.0 & 0.10 \\
192.69 & 0.64 & 0.37 & 0.18 \\
193.27 & 0.58 & 0.25 & 0.16 \\
193.89 & 0.62 & 0.26 & 0.23 \\
\hline
\multicolumn{4}{||c||}{V$\rm _{ion}$ = 0.48 kV; N ions only; deposition rate
0.4 \AA/s;
approx. 15\% $sp^3$ and 85\% $sp^2$} \\
\hline
191.74 & NA & 1.0 & 0.10 \\
192.39 & 0.65 & 0.31 & 0.18 \\
192.98 & 0.59 & 0.16 & 0.16 \\
193.6 & 0.63 & 0.21 & 0.16 \\
\hline
\multicolumn{4}{||c||}{V$\rm _{ion}$ = 1 kV; N ions only; deposition rate 1.7
\AA/s;
approx. 100\% $sp^2$} \\
\hline
191.93 & NA & 1.0 & 0.52 \\
192.69 & 0.76 & 0.19 & 0.17 \\
193.19 & 0.5 & 0.18 & 0.27 \\
194.0 & 0.81 & 0.22 & 0.37 \\
\hline
\multicolumn{4}{||c||}{hBN powder}\\
\hline
192.0 & NA & 1.0 & 0.04\\
\hline
\multicolumn{4}{||c||}{rBN powder}\\
\hline
192.26& NA & 1.0 & 0.10\\
192.92& 0.659 & 0.30 & 0.02\\
193.52& 0.60 & 0.27 & 0.02 \\
194.3& 0.78 & 0.34& 0.1 \\
\hline
\hline
\end{tabular}
\end{center}
\bigskip
\caption{Parameters obtained from fits to the B K-edge
spectra of two types of BN powder and three
BN/Si films prepared by ion-assisted pulsed laser deposition under
slightly different conditions.   The
$sp^3$ fraction in the films was estimated using infrared transmission
measurements.  NA = not applicable.}
\end{table}
\end{small}
}

B and N K-edge x-ray absorption spectra were obtained at both the 8-1
\pagebreak
and 8-2 beamlines at SSRL, as well as at the IBM/U8 beamline at NSLS.
Details of the data acquisition have been reported
previously.\cite{terminello93,BNapl}
A simple polynomial was adequate to fit the background of the powder
spectra but it was necessary to use a gaussian-broadened step
function\cite{stohrbook} to fit the background intensity for all the
films' B K-edge spectra.  The position of the step was fixed at 2.125 eV
above the B 1s $\pi^*$ peak, the energy where $\sigma ^*$ absorption
begins in cBN.  The absolute energies of spectral features are
reproducible to within 0.3 eV from run to run, but the relative
energies of the features are reproducible to within 0.05 eV.

\begin{flushleft}
{\bf RESULTS}
\end{flushleft}

Figure~\ref{b1sspectra}a shows the B K-edge spectra of the three BN/Si
films, while Figure~\ref{b1sspectra}b shows the spectra of hBN, rBN
and cBN powders, which have been described in detail
previously.\cite{terminello93} The most notable feature of the hBN and
rBN powder data is the presence of the $\pi^*$ feature at 192.0 eV,
which is characteristic of $sp^2$ bonding.\cite{fomichev68,davies81}
This $\pi^*$ feature is
absent in cBN powder due to the $sp^3$ nature of the bonding.  In the
$sp^2$-bonded layered materials, the $\pi^*$ feature has been
\pagebreak
described as a core exciton whose position below the conduction-band
edge is a result of a 1.3 eV excitonic binding energy.\cite{carson87}
Comparison of the data
Figures~\ref{b1sspectra}a and \ref{b1sspectra}b shows that the bonding
in the BN/Si films is predominantly $sp^2$, similar to the hBN powder.
However, there are three sharp peaks that appear in the B spectra of
all three BN films just above the $\pi^*$ peak but which don't appear
in the hBN data.   Parameters obtained from fits to these smaller
peaks are collected in Table 1 along with fit parameters from the hBN
and rBN powders.  The BN/Si films' spectra also have a peak at 199 eV in the
middle of the empty $\sigma$ band which is not present in any of the
powder spectra.

\begin{figure}
\vspace{18cm}
\vfill
\caption[]{a) B K-edge x-ray absorption spectra for three BN/Si films.
The x-ray flux was directed onto the films at normal incidence.
Notice the 3 small peaks between the $\pi*$ peak at 192.0 eV and the
onset of the $\sigma^*$ band at 197 eV.
b) B K-edge spectra for hBN, cBN, and rBN powders.}
\label{b1sspectra}
\end{figure}

N K-edge spectra for the BN/Si films and the three BN powders are
shown in Figure~\ref{n1sspectra}.  The first peak at 402 eV in the
$sp^2$-bonded materials is again a $\pi^*$ feature, although it is not
as intense or as far below the conduction-band edge as in the B K-edge
spectra.  In the Wannier model of excitonic behavior, the intensity
and binding energy of an exciton are  reduced when the
core hole is created on the anionic site.\cite{carson87} Overall the
film data is similar to that for hBN and rBN, although the intensity
of the $\pi^*$ peak at 402 eV is reduced with respect to that of the
the $\sigma^*$ peak at 408.5 eV in the films compared to the hBN or
rBN standards.

\begin{figure}
\vspace{18cm}
\vfill
\caption[]{a) N K-edge x-ray absorption spectra for three BN/Si films.
 All data were taken with x-rays at normal incidence.
b) N K-edge spectra for hBN, cBN, and rBN powders.}
\label{n1sspectra}
\end{figure}

\begin{flushleft}
{\bf DISCUSSION}
\end{flushleft}

Because infrared spectroscopy on these BN films has indicated the
presence of cBN, it is logical to ask whether the unexplained features
in the films' spectra may result from a combination of hBN and cBN
states.  In fact, the occurrence of a step near 194 ev in the
background intensity of the films' B K-edge spectra is probably an
indication of the cBN phase.  The peak at 199 eV in the BN/Si B K-edge
data may also be due to a contribution from cBN.  Also attributable to
cBN is the increased intensity of the $\sigma^*$ peak with respect to
the $\pi^*$ peak in the films N K-edge spectra.  The size of the
above-mentioned deviations from the hBN spectra appears to be
correlated with the amount of cBN indicated in each film by the IR
measurements.

The three fairly narrow peaks near 192.6, 193.2, and 194.0 eV in the
films' B K-edge data (see Table 1) do not have analogs in the cBN or
hBN data, nor are there similar features in the K-edge spectra of
disordered C films.\cite{denley80}  With much lower energy
resolution, Fomichev and Rumsh previously reported a single broad peak
near 194.0 eV in their study of hBN powder.\cite{fomichev68} The
nearly constant energy spacing shows that these peaks are not a
Rydberg series\cite{stohrbook} and the approximately 0.6 eV magnitude
of the spacing rules out an origin involving lattice vibrations.  The
proximity of these sharp peaks to the $\pi^*$ excitonic peak suggests
a related origin.  One possibility is that stacking
in the films is disordered and that the individual peaks represent
different stacking configurations.
Given that BN films are often
non-stoichiometric,\cite{wada92} another possibility is that these
peaks represent a quasi-bound final state formed by a photoelectron
and a N vacancy.  The most
obvious possibility is that these peaks are due to the presence of
rBN, whose B K-edge spectrum shows similar small peaks, as documented
in Table 1.  The unidentified peaks would then be due to modification
of the $\pi$ bonding by the ABCA stacking that is characteristic of
rBN.  Positive identification rBN
as a minor constituent of these films would require further study.

A closer look at Table 1 and Figures~\ref{b1sspectra}a and
\ref{n1sspectra}a shows that the peaks for the film synthesized with
the higher ion beam voltage and higher deposition rate are
considerably broader than those for the other two films.  Energetic
ions have likely caused damage to this film, corroborating the
findings of a previous IR and photoemission spectroscopy study by Wada
and Yamashita on ion-assisted evaporation of BN films.\cite{wada92}
Wada and Yamashita also observed metallic B and a smaller amount of
\pagebreak
cBN in material grown at deposition rates above 1 \AA /s.\cite{wada92}
These results are consistent with the observation (see Table 1) that the film
synthesized at the higher deposition rate has a lower fraction of $sp
^3$-bonded material.  Given the evident degree of disorder in this
film, the recent observation of a preferred crystallographic
orientation came as a surprise.    NEXAFS and transmission electron
microscopy experiments showed   the hexagonal layer planes were oriented
close to orthogonal to the substrate.\cite{BNapl}  This
orientation is contrary to that usually observed in thin graphite
films, where the c-axis is typically normal to the film-substrate
interface.\cite{denley80}

\begin{flushleft}
{\bf CONCLUSIONS}
\end{flushleft}

B and N K-edge spectra have been acquired on both single-phase BN
powders and on BN/Si films fabricated using ion-assisted pulsed-laser
deposition.  Comparison with the powder data shows the films to be
primarily $sp^2$-bonded.  Systematic differences between the films'
spectra and that of pure hBN suggest that another phase is present,
probably the cBN that has been observed with infrared transmission
measurements.  Three sharp peaks appear in the films' B K-edge spectra
between the usual $\pi^*$ and $\sigma^*$ bands.  These peaks may
be related to disorder or may indicate the presence of the rBN phase
in the films.  Angle-dependent x-ray absorption studies have shown
that the BN/Si films are preferentially oriented with the hexagonal
axis nearly in the film plane.  Transmission electron microscopy and
photoemission studies are underway to help clarify issues of film
morphology and stoichiometry.   The near-edge x-ray absorption spectra
of thin BN films are proving to be more complex to interpret than those of
vapor-deposited C films.

We would like to thank A.K. Ballal and L. Salamanca Riba for
the TEM pictures, D.K. Shuh for assistance with the data collection
and E.A. Hudson for helpful discussions. Part of this work was
performed under the auspices of the U.S. Department of Energy by LLNL
under contract No. W-7405-ENG-48.

\vspace{-0.7cm}
\begin{small}

\end{small}


\begin{thebibliography}{99}

\bibitem{stohrbook}
J. St\"ohr, \underline{NEXAFS Spectroscopy}, (Springer Verlag, New
York, 1992).

\bibitem{rosenberg86}
R.A. Rosenberg, P.J. Love, and V. Rehn, Phys. Rev. {\bf B 33}, 4034 (1986).

\bibitem{morar85}
J.F. Morar, F.H. Himpsel, G. Hollinger, G. Hughes, and J.L. Jordan,
Phys. Rev. Lett. {\bf 54}, 1960 (1985).

\bibitem{denley80}
D. Denley, P. Perfetti, R.S. Williams, D.A. Shirley, and J. St\"ohr,
Phys. Rev. {\bf B21}, 2267 (1980).

\bibitem{terminello93}
L.J. Terminello, A. Chaiken, D.A. Lapiano-Smith, G.L. Doll and T.
Sato, submitted to J. Vac. Sci. Techn.

\bibitem{ballal92}
A.K. Ballal, L. Salamanca Riba, G.L. Doll, C.A. Taylor, and R. Clarke,
J. Mater. Res. {\bf 7}, 1618 (1992).

\bibitem{BNapl}
A. Chaiken, L.J. Terminello, J. Wong, G.L. Doll and C.A. Taylor II,
Appl. Phys. Lett. {\bf 63}, 2112 (1993).

\bibitem{fomichev68}
V.A. Fomichev and M.A. Rumsh, J. Phys. Chem. Solids {\bf 29}, 1015 (1968).

\bibitem{davies81}
B.M. Davies, F. Bassani, F.C. Brown, and C.G. Olson, Phys. Rev. {\bf
B24}, 3537 (1981).

\bibitem{carson87}
R.D. Carson and S.E. Schnatterly, Phys. Rev. Lett. {\bf 59}, 319 (1987).

\bibitem{wada92}
T. Wada and N. Yamashita, J. Vac. Sci. Techn. {\bf A10}, 515 (1992).

\end{thebibliography}
\end{document}